\documentclass[a4paper]{spie}  

\usepackage{amsmath,amsfonts,amssymb}
\usepackage{graphicx}
\usepackage[colorlinks=true, allcolors=blue]{hyperref}
\usepackage{makecell}
\usepackage[T1]{fontenc}

\usepackage{enumitem}
\usepackage{float}
\usepackage{multirow}

\title{VERITAS 2.3.1: Optimisation and Characterisation of the Enhanced Readout ASIC for the NewAthena Wide Field Imager}

\author[a]{Ajay~Kumar~Dakshinamurthy*}
\author[b]{Sven Herrmann}
\author[b]{Peter Orel}
\author[a]{Anna-Katharina~Schweingruber}
\author[a]{Astrid Mayr} 
\author[a]{Johannes~Müller-Seidlitz} 
\author[a]{Jonas Reiffers} 
\author[a]{Sebastian Albrecht} 
\author[b]{Steven W. Allen}
\author[b]{Glenn Morris}

\affil[a]{Max-Planck Institute for Extraterrestrial Physics, Giessenbachstr. 1, 85748 Garching, Germany}
\affil[b]{Kavli Institute for Particle Astrophysics and Cosmology, Stanford University, Stanford, CA, USA}

\authorinfo{*ajaykd@mpe.mpg.de, www.mpe.mpg.de}

\begin{document} 
\maketitle

\begin{abstract}
VERITAS 2.3.1 is the next iteration of the VErsatile Readout based on Integrated Trapezoidal Analogue Shapers (VERITAS) integrated circuit (IC) architecture for high-speed, low-noise readout of DEPleted Field Effect Transistor (DEPFET) detectors in the Wide Field Imager (WFI) on ESA’s NewAthena X-ray satellite. Building on VERITAS 2.3, which demonstrated a short processing time of 2.5 $\mu$s per readout and a system noise target of about $3\,\mathrm{e^-}$ ENC RMS, the VERITAS 2.3.1 revision has been improved with additional features and targeted optimisations of existing analogue and digital blocks, alongside refined layout routing to reduce parasitic effects, to improve transient behavior, cross talk, manufacturability, and reliability while preserving the proven VERITAS architecture. The functionality and performance of VERITAS 2.3.1 was characterised in two steps. First by using a dedicated application-specific integrated circuit (ASIC) only test setup. Second, a full-scale module test setup is used, with integrated DEPFET sensors and readout electronics, under vacuum and at mission-like temperatures. This paper presents the design updates and implementation details of the VERITAS 2.3.1, compares its measured performance to that of VERITAS 2.3, and discusses the impact of the added features, block-level optimisations, and routing improvements on overall system performance. 
\end{abstract}


\section{INTRODUCTION}
\label{sec:intro}  
NewAthena (New Advanced Telescope for High-ENergy Astrophysics) is ESA’s next-generation X-ray observatory and the successor to the Athena mission concept developed under the Cosmic Vision 2015–2025 programme. The mission is designed to investigate some of the most energetic phenomena in the Universe, including black holes and the formation and evolution of large-scale cosmic structures. NewAthena will carry two primary scientific instruments: the X-ray Integral Field Unit (X-IFU) and the Wide Field Imager (WFI). Following the endorsement of a rescoped mission concept by ESA’s Science Programme Committee (SPC) in November 2023, NewAthena is currently in the study phase. Subject to successful completion of the mission design and cost assessment, mission adoption is planned for 2027, with launch anticipated in the late 2037.\cite{nandra2018athena}

The WFI provides Fano-limited spectroscopic performance over an energy range of ~0.2 keV to 15 keV, combined with a large field of view (40$\,$'\,x\,40$\,$') and high count-rate capability. The detector system is based on multiple arrays of DEPFET sensors and comprises two detector types with different pixel configurations. The Large Detector Array (LDA) consists of four Large Detectors (LDs), each with 512\,x\,512 pixels, while the Fast Detector (FD) comprises a single 64\,x\,64 pixel matrix. The LDA is optimised for wide-field observations, whereas the FD is designed for high-count-rate measurements of bright point-like sources.

The detector modules operate in rolling-shutter mode, with front-end electronics comprised of two arrays of ASICs, each dedicated to their specific tasks. The Switcher ASIC sequentially activates up to 64 DEPFET sensor rows, while the VERITAS ASIC simultaneously reads out these 64 pixels from the selected row. Consequently, each LD requires eight Switcher and eight VERITAS ASICs, respectively. The FD employs a split-frame readout architecture, dividing the detector into two halves and enabling the simultaneous activation and readout of two rows using two Switcher and two VERITAS ASICs.\cite{meidinger2020development} The VERITAS 2.3.1 ASIC is designed to accommodate different readout schemes. That is, it can read out DEPFETs and other Charge coupled devices (CCD) in the so called source follower mode, as well as DEPFETs in the drain current mode. The latter has been chosen as the primary readout mode of operation for the WFI detector array. This paper will focus solely on the drain readout mode of operation, which delivers a readout time of 2.5\,µs while maintaining an excellent noise performance of $\sim$ 3\,\(e^-\) ENC RMS, leading to a frame rate of \( \leq \)\,5\,ms/frame for the LDA and \( \leq \)\,80\,µs/frame for the FD, respectively.

The WFI signal chain consist of the VERITAS 2.3.1 ASIC, which reads out up to 64 detector channels simultaneously. The ASIC also performs signal conditioning in the form of correlated double sampling method, which completely done in analog. The resulting signal is then sampled and multiplexed onto a single analog output that feeds into an ADC. This results in eight analogue output signals from the ASICs, one for each of the LDs and two more from the ASICs of the FD. These signals are then digitised by an analogue-to-digital converter (ADC) array inside the Detector Electronics Sub-system (DES). There are five DES units - one for each of the LDs and an additional one for the FD. The DES also provides power supply and sequencer signals to control, clock, and program the ASICs.\cite{astrid2024}

\begin{figure} [ht]
    \begin{center}
    \begin{tabular}{c} 
    \includegraphics[width=0.6\textwidth]{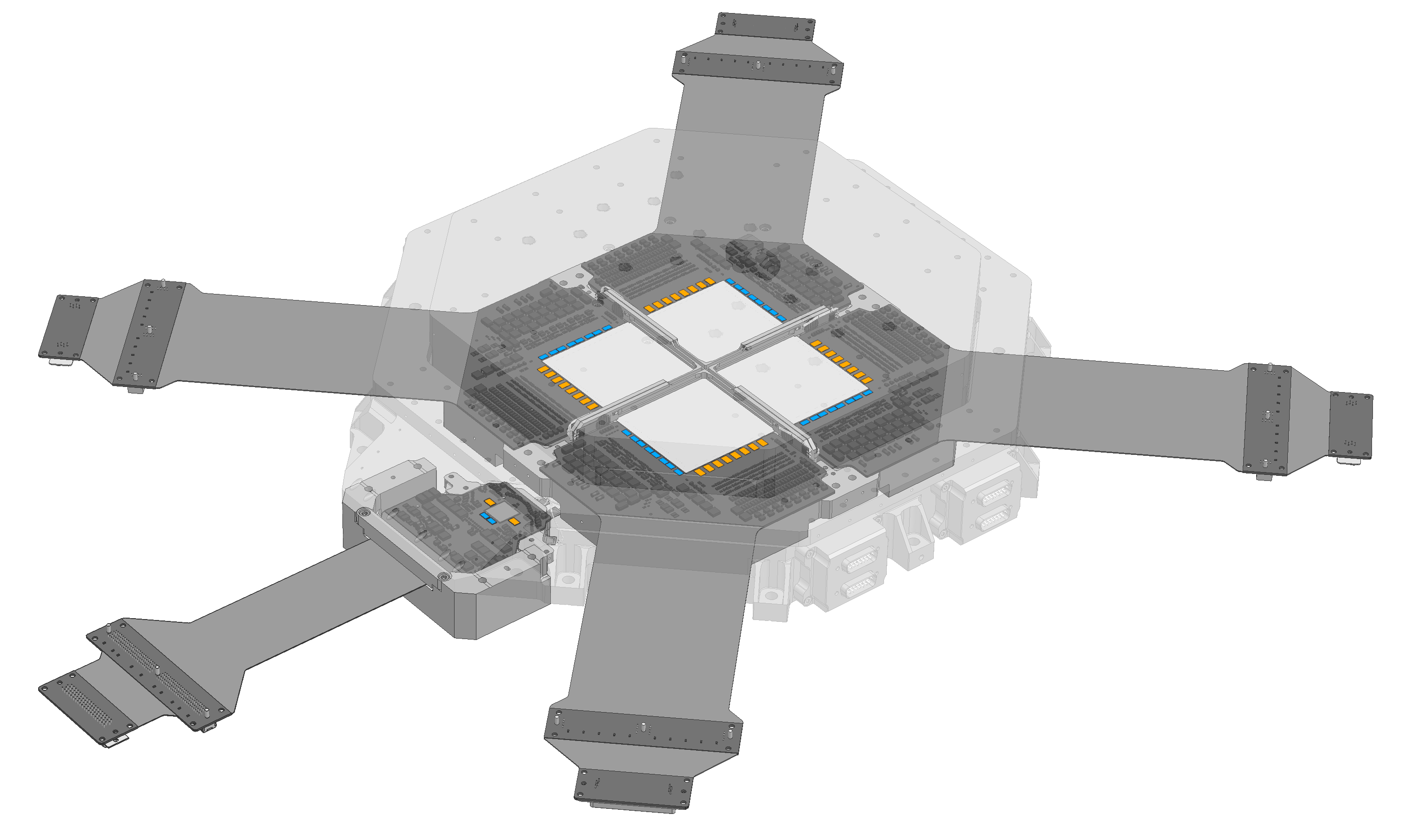}
    \end{tabular}
    \end{center}
   \caption[example] 
   { \label{fig:WFI} 
    Model of the camera head: in the centre, 4 Large Detectors (LD) building the Large Detector Array (LDA) with each 8 VERITAS 2.3.1 in orange and 8 SwitcherA ASICs in blue, and in the lower left corner, the Fast Detector (FD) with 2 VERTIAS 2.3.1 in orange and 2 SwitcherA ASICs in blue. Flex leads that connect to DES can also be seen in the figure.}
\end{figure} 

\section{VERITAS-Based Readout of DEPFET Drain Current Signals} 
The DEPFET consists of a PMOS transistor fabricated on a sidewards-depleted n-type substrate. Appropriate biasing of the device terminals creates a potential minimum beneath the transistor channel, referred to as the internal gate, where electrons generated by incident X-rays are collected. The accumulated charge modulates channel conductivity, resulting in a drain current proportional to the collected signal charge.\cite{kemmer1987new} Two readout schemes are commonly employed: source follower and drain current readout. For the WFI, drain current readout is selected due to its faster operation, enabled by the absence of RC settling limitations.

\begin{figure} [H]
    \begin{center}
    \begin{tabular}{c} 
    \includegraphics[width=0.7\textwidth]{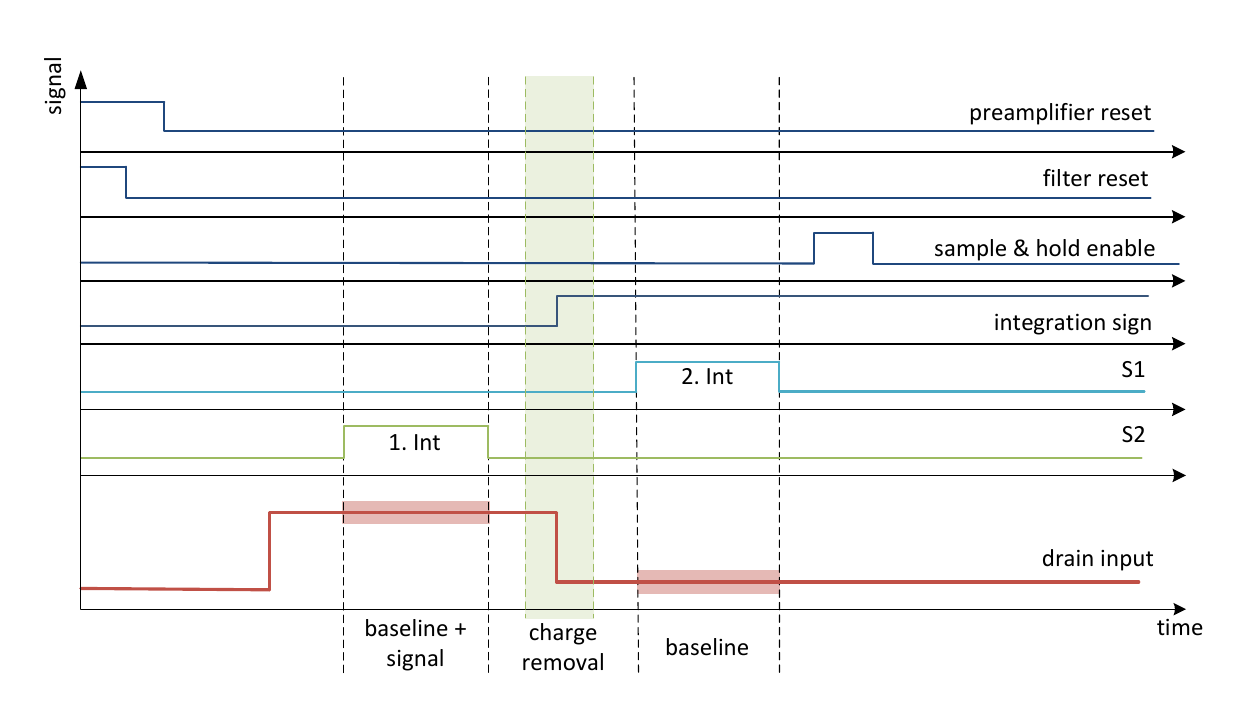}
    \end{tabular}
    \end{center}
    \caption[example] 
    { \label{fig:Seq} 
    The time diagram shows the sequential steps in the signal processing cycle for obtaining the DEPFET signal by performing two integrations: signal plus baseline and baseline. It highlights the integration phases, signal-processing intervals, and control signals involved in processing incident signals through the analogue channel stages of VERITAS\,2.3.1.}
\end{figure}

Charge removal, as shown in figure~\ref{fig:Seq}, is performed by the clear-FET, which transfers the collected electrons from the internal gate to the clear node when appropriate voltages are applied. This process resets the DEPFET for the next acquisition cycle. The collected charge is determined by subtracting a baseline measurement, acquired after clearing, from a signal-plus-baseline measurement acquired prior to the clear operation.\cite{meidinger2020development}

\begin{figure} [ht]
    \begin{center}
    \begin{tabular}{c} 
    \includegraphics[width=0.7\textwidth]{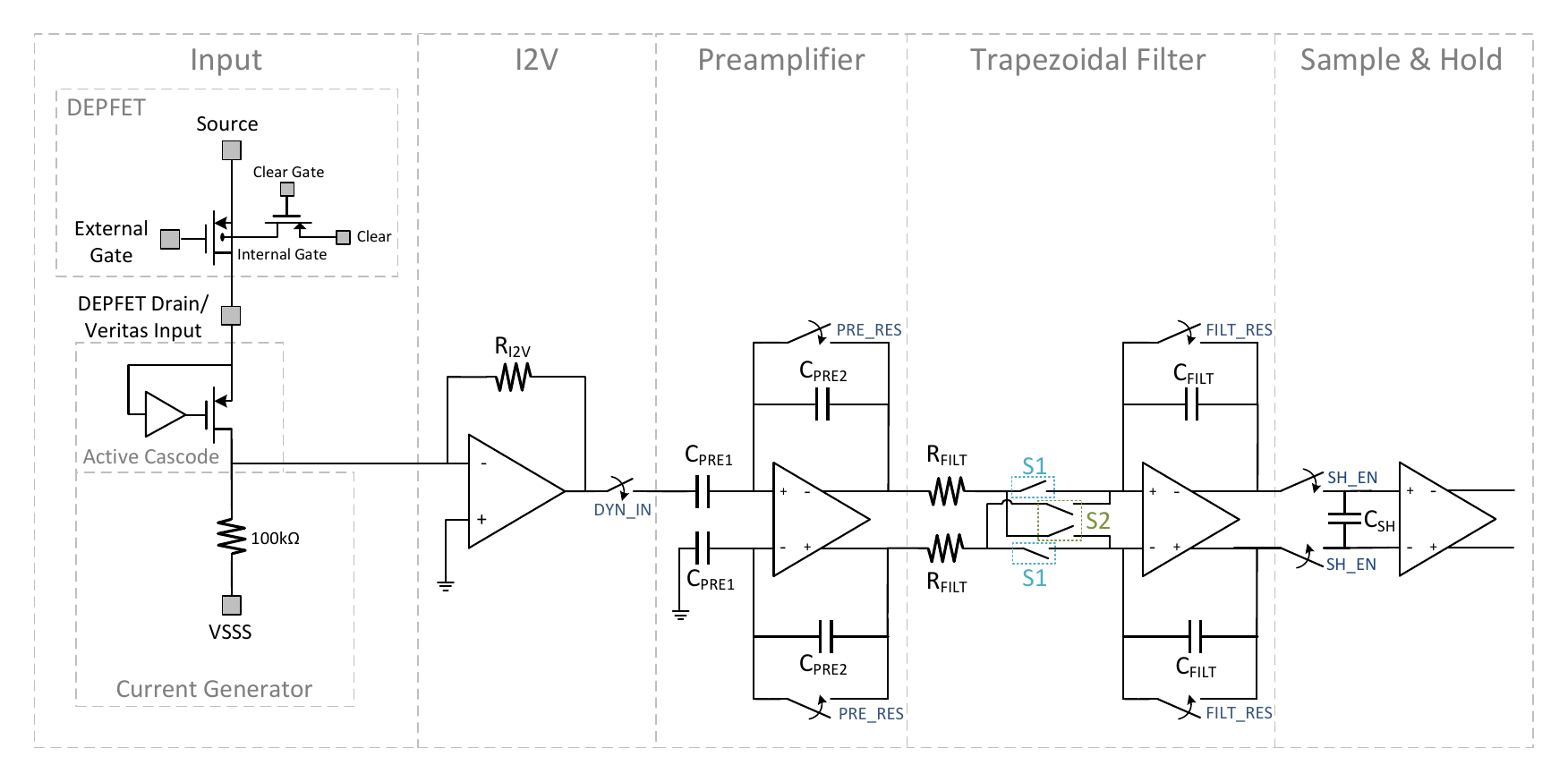}
    \end{tabular}
    \end{center}
    \caption[example] 
    { \label{fig:Channel} 
    Overview schematic of one DEPFET pixel connected to a single analogue channel within the VERITAS\,2.3.1, including the main stages, components, and switches involved in the signal processing cycle.}
\end{figure}

In drain current readout mode, the DEPFET drain is directly connected to the VERITAS input, as shown in figure~\ref{fig:Channel}. A programmable current source within the VERITAS biases the DEPFET. The detector output current consists of a large DC bias component and a smaller signal component proportional to the collected charge. The current generator subtracts the nominal DC current, leaving a signal current that is processed by the current-to-voltage (I2V) stage.

An active cascode (AC) is placed between the DEPFET and the I2V input to isolate the sensitive I2V input node from the highly capacitive column line formed by the interconnected DEPFET drains, allowing the I2V amplifier to maintain stability and providing a stable voltage to the DEPFET drain. Following the I2V conversion, the signal is AC-coupled and further amplified by a fully differential preamplifier, which has a capacitive feedback and provides user selectable voltage gain.

The filter stage performs the correlated double sampling method by using a fully differential amplifier configured as an integrator with switched directions and pass though modes. More accurately, two sequential integrations are performed: one on the signal plus baseline and one on the baseline alone but in the opposite direction, thus subtracting the baseline. The DEPFET signal is then stored in the sample and hold stage.\cite{herrmann2018veritas, porro2014veritas, porro2013veritas, porro2010asteroid}

\begin{figure} [ht]
\begin{center}
\begin{tabular}{c} 
\includegraphics[width=0.9\textwidth]{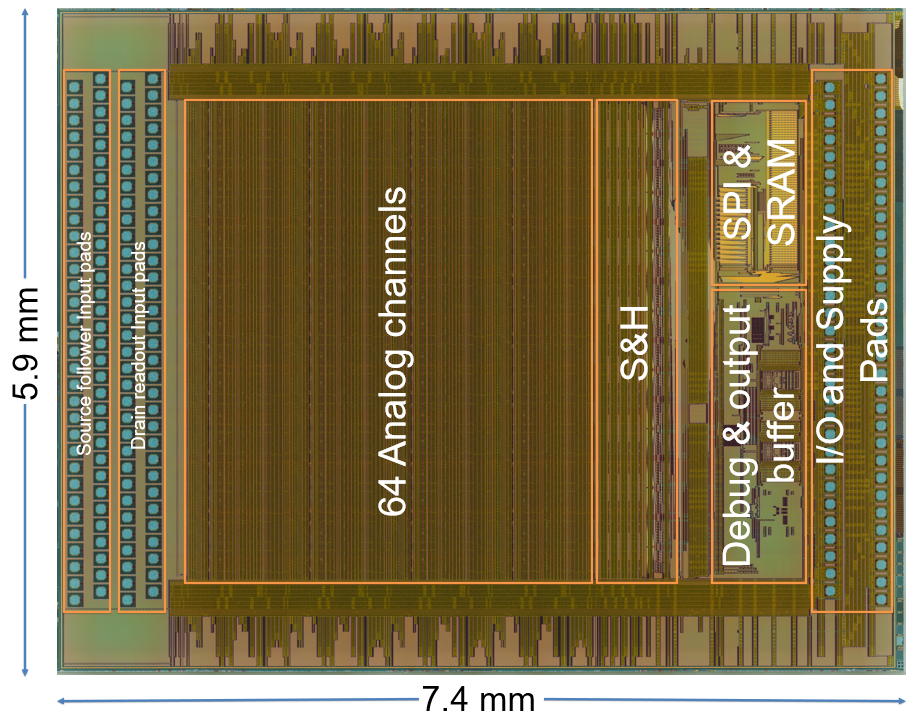}
\end{tabular}
\end{center}
\caption[example] 
{ \label{fig:die} 
Photograph of the fabricated VERITAS\,2.3.1 die, which measures 7.4\, mm x 5.9\, mm. On the left side, there are two sets of 64 inputs for the analogue channels, which can be used for either source follower or drain current readout. On the right side, there are 61 pads for power supply, analogue I/O, digital I/O, and analogue monitoring. The middle part comprises the 64 analogue channels, covered with supply lines, the multiplexer, the digital part, and the output buffer.}
\end{figure}

\section{OPTIMISATIONS}
\subsection{Output buffer and offset voltage shift (OVS)}

The output buffer is implemented as a fully differential unity-gain amplifier (FDA) incorporating both Common-Mode Feedback (CMFB) and Offset Voltage Shift (OVS) functionality. The CMFB circuit continuously monitors the output common-mode voltage and actively regulates it, ensuring stable operation and maintaining the desired operating point under varying signal conditions.

The OVS function is controlled by one of the seven on-chip digital-to-analogue converters (DACs) and provides a programmable DC offset control of the buffer output voltage. This enables the VERITAS output signal range to be adjusted and optimally matched to the input range of the external analogue-to-digital converter (ADC), thereby maximising the available dynamic range.

\begin{figure}[H]
\centering

\begin{minipage}{0.4\textwidth}
    \centering
    \includegraphics[width=\textwidth]{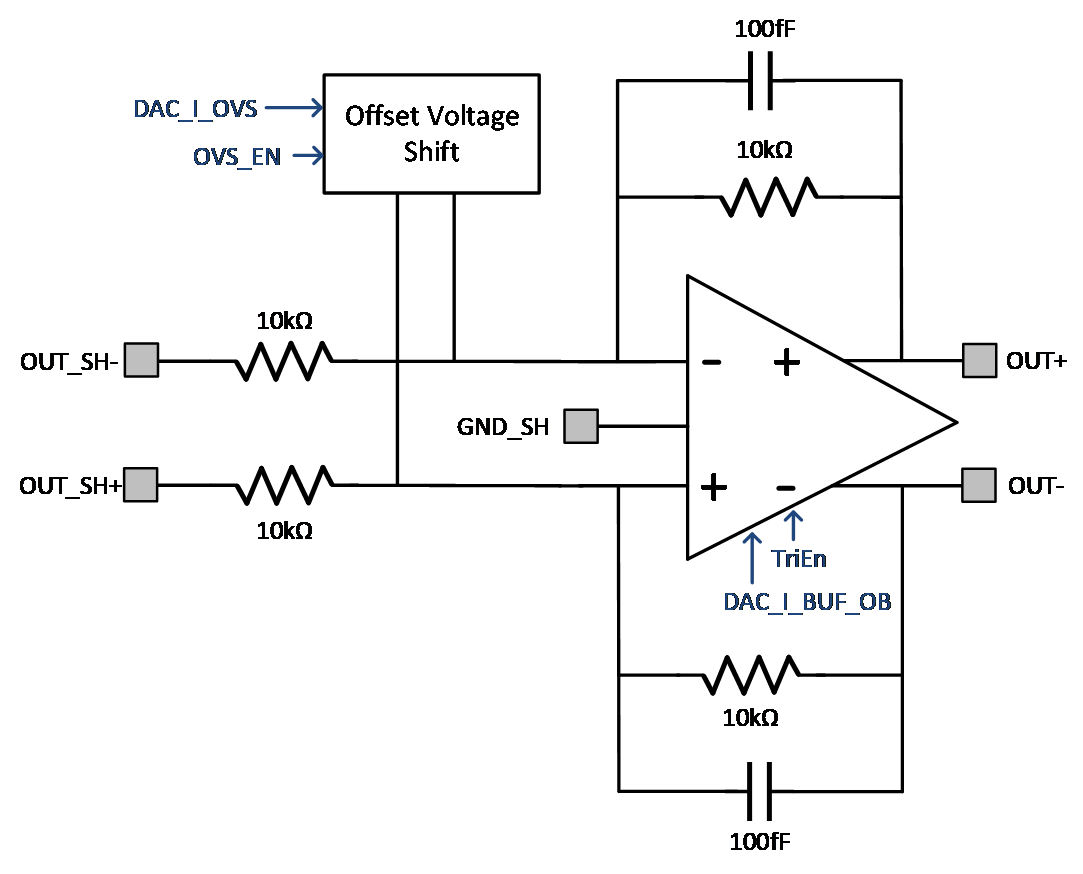}
   
\end{minipage}
\hfill
\begin{minipage}{0.55\textwidth}
    \centering
    \includegraphics[width=\textwidth]{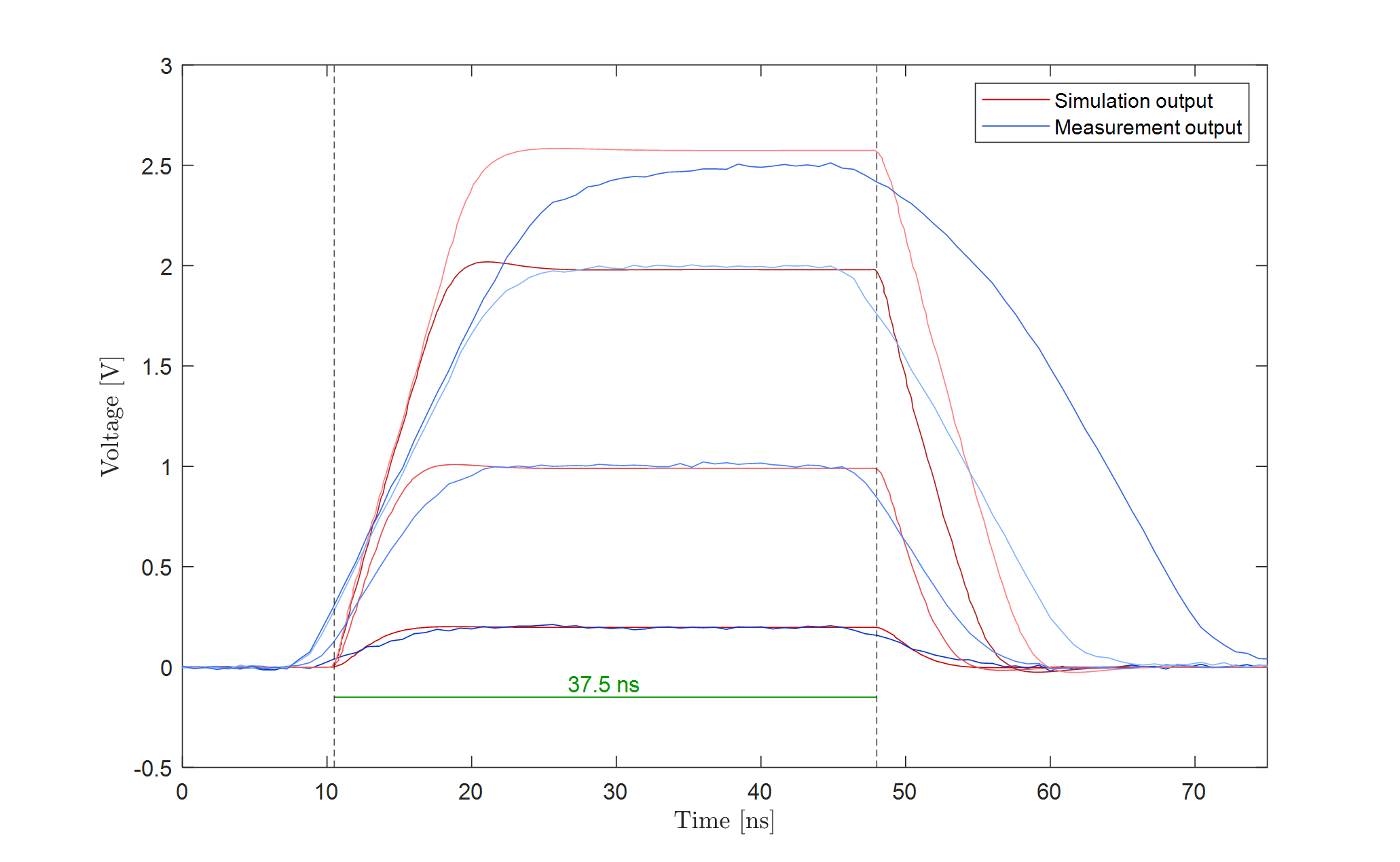}
  
\end{minipage}

\caption[example]
{
\label{fig:OB}
Comparison of the simulated and measured transient responses of the output buffer under output voltage shift (OVS) conditions.}
\end{figure}

The output signal is digitised by an external 14-bit radiation-hardened Analogue Devices AD9246S ADC located on the Frame Processing Module of the Detector Electronics. Operating at a sampling rate of 80 MHz, the ADC provides accurate digitisation of the VERITAS output signal. To minimise parasitic effects, noise coupling, and signal degradation, the ASIC must be positioned in close proximity to the ADC. Figure~\ref{fig:OB} presents the schematic of the output buffer, including the feedback network and the offset-current injection circuitry used to realise the OVS functionality.

\subsection{Reference generator}

In VERITAS 2.3~\cite{herrmann2018veritas,anna2024}, the reference generator exhibited premature saturation, reaching a maximum output current of approximately $135\,\mu\mathrm{A}$ at $V_{\mathrm{REF}} = -250\,\mathrm{mV}$. This corresponds to a current-programming resolution of approximately $1.06\,\mu\mathrm{A}$ per DAC least-significant bit (LSB), and consequently limited the achievable bias-current range. As a result, the circuit did not meet the design specification, which required an output current of $200\,\mu\mathrm{A}$ at $V_{\mathrm{REF}} = -500\,\mathrm{mV}$. This target corresponds to a current-programming resolution of approximately $1.57\,\mu\mathrm{A}$ per DAC LSB. In VERITAS 2.3.1, this limitation was resolved by modifying the reference-generator circuitry. Subsequent electrical characterisation verified the extended operating range and improved linearity, demonstrating compliance with the target current and voltage specifications across the full DAC control range.

\subsection{Improved reliability}
To improve power-supply integrity and overall circuit reliability, the resistance of the on-chip power-distribution network was reduced through layout optimisation. Lower routing resistance minimises voltage drops along the supply rails caused by current flow, thereby reducing IR drop across the ASIC. Design targets were established to limit the resistance of the primary $V_{\mathrm{DD}}$ and $V_{\mathrm{SS}}$ supply networks to less than $1\,\Omega$, while maintaining the resistance of all other analogue and digital supply rails below $3\,\Omega$. Post-layout extraction and verification confirmed that these targets were successfully achieved across the entire power-distribution network. As a result, more stable supply voltages are delivered to critical analogue and digital circuit blocks, reducing sensitivity to process, voltage, and temperature variations. The improved supply regulation enhances bias stability, operating-point accuracy, and signal integrity, particularly under high-current operating conditions. Consequently, the ASIC exhibits increased robustness and more predictable performance across its intended operating range.

\subsection{Power saving in Debugging buffer}
In previous revisions, the debugging buffer remained partially active when not in use, with both the biasing circuitry and output stage continuing to draw static current. This unnecessary power consumption contributed to increased residual idle power and reduced overall power efficiency. To address this issue, the debugging buffer architecture was redesigned to incorporate a dedicated disable functionality. When the buffer is not required for operation or characterisation, the biasing network and output stage can now be completely switched off, significantly reducing residual idle and standby power consumption. This optimisation improves overall power efficiency while maintaining full functionality when the debugging buffer is enabled.

\subsection{Parasitics reduction}
The layout of VERITAS 2.3.1 was extensively refined to minimise parasitic effects and improve overall signal integrity. Careful optimisation of critical routing paths was performed to reduce parasitic resistance and capacitance, which can significantly degrade analogue performance, particularly in low-noise, high-speed readout circuits. Excessive parasitics can adversely affect gain accuracy, settling behaviour, noise performance, bandwidth, and channel-to-channel matching, making layout optimisation a key factor in achieving the targeted system performance.

Particular attention was given to the signal paths between the filter stage and the sample and hold (S\&H) circuitry. In VERITAS 2.3, the routing arrangement was susceptible to crosstalk between the differential filter outputs and the differential S\&H inputs, potentially introducing unwanted signal coupling and degrading measurement accuracy. The routing topology was therefore redesigned to improve isolation and reduce coupling between adjacent signal lines.

Further optimisation was applied to the drain current readout input routing to minimise the series resistance. This reduced the AC path impedance, resulting in a more stable and uniform voltage distribution across all DEPFET channels. Consequently, voltage drops at this highly sensitive analogue input node were minimised, improving signal integrity and readout fidelity.\cite{herrmann2018veritas, anna2024} In addition, connection between the S\&H output and the input of the output buffer. To ensure a negligible impact on the closed-loop gain of the output buffer, the routing resistance was minimised and maintained well below 1 $\%$ of the buffer input resistance.

\section{CHARACTERISATION} 
\begin{figure} [ht]
    \begin{center}
    \begin{tabular}{c} 
    \includegraphics[width=\textwidth]{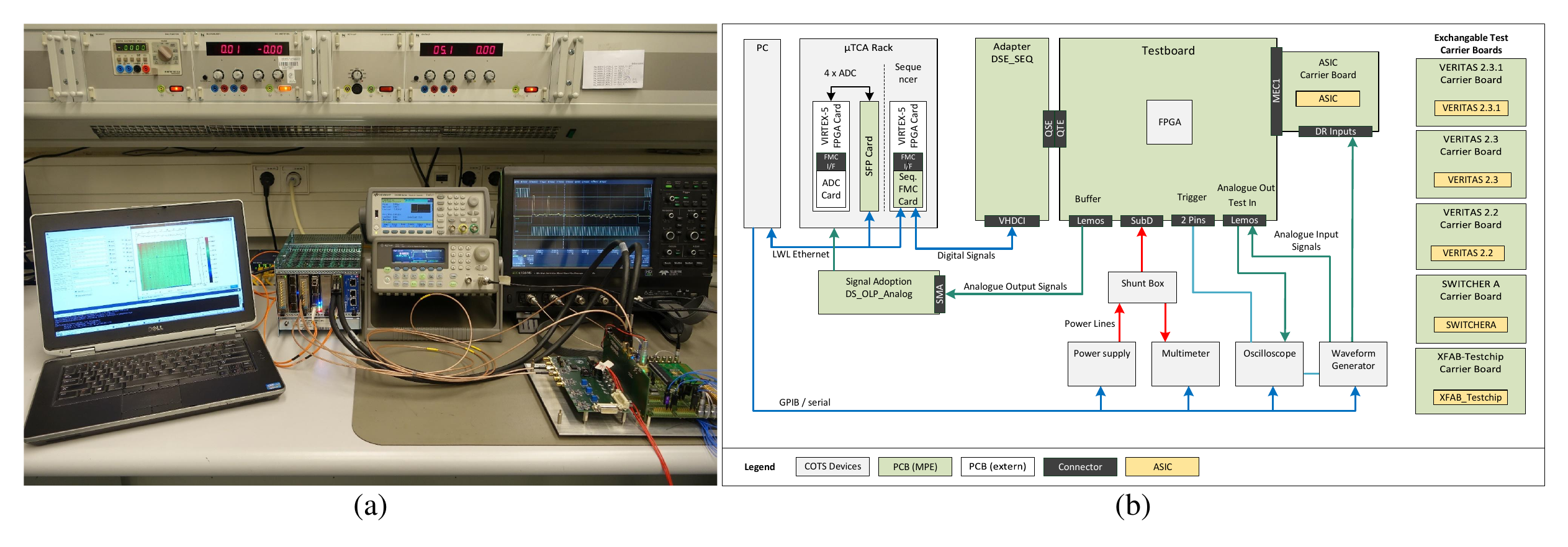}
    \end{tabular}
    \end{center}
    \caption[example] 
        { \label{fig:ASICsetup} 
        Modular ASIC test setup for standalone testing of ASICs. (a) Photograph (b) Block diagram}
\end{figure}

VERITAS 2.3.1 underwent comprehensive functionality and performance characterisation using a dedicated test setup developed and continuously refined at the Max Planck Institute for Extraterrestrial Physics (MPE). The modular ASIC test setup is specifically designed to support standalone evaluation of a range of front-end ASICs, including both VERITAS and Switcher devices. Such testing is essential to verify correct operation and compliance with performance requirements prior to integration with detector systems in laboratory environments. The modular architecture of the test platform shown in Figure~\ref{fig:ASICsetup} provides high flexibility, enabling rapid reconfiguration to accommodate different ASIC variants and measurement requirements.\cite{herrmann2018veritas, anna2024}

\begin{figure} [ht]
    \begin{center}
    \begin{tabular}{c} 
    \includegraphics[width=0.7\textwidth]{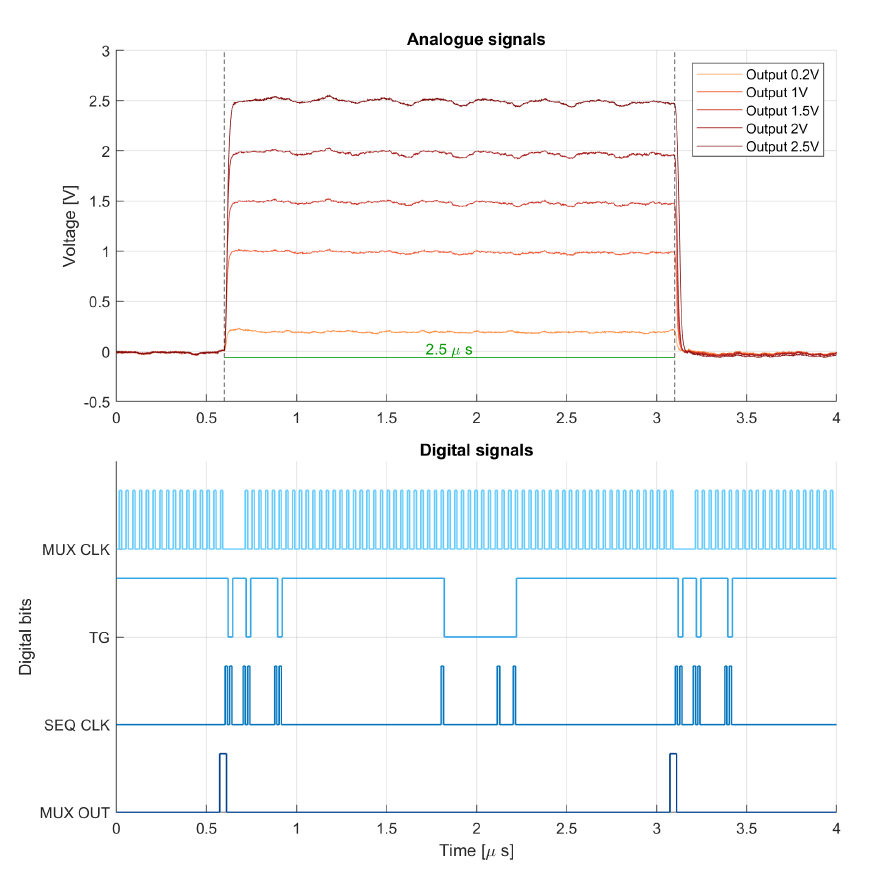}
    \end{tabular}
    \end{center}
   \caption[example] 
   { \label{fig:64channel} 
    The measured response across all 64 output buffer channels, together with the corresponding digital outputs, exhibits excellent uniformity and confirms the correct functionality of VERITAS 2.3.1, providing validation of the device prior to comprehensive electrical and spectroscopic characterisation.}
\end{figure}

\subsection{I2V feedback resistance, dynamic range, linearity, and noise}
A series of measurements was conducted to evaluate the feedback resistance, dynamic range, linearity, and noise performance of the I2V converter for each programmable gain setting. For these measurements, a voltage pulse generated by a waveform generator was injected into the VERITAS 2.3.1 input through a precision series resistor, thereby emulating the detector current signal. The pulse baseline was adjusted according to the bias current (I$_\text{BIAS}$) of the internal current source to ensure proper cancellation of the nominal drain current (I$_\text{DRAIN}$). As a result, only the pulse amplitude contributed to the effective input current (I$_\text{INPUT}$) presented to the I2V stage. This approach enabled controlled characterisation of the converter performance over its full operating range.\cite{porro2014veritas, herrmann2018veritas, anna2024} \\

\begin{table}[ht]
\caption{Results of I2V feedback resistance measurement.} 
\label{tab:I2v_table}
\begin{center}       
\begin{tabular}{|c|c|c|c|c|} 
\hline
\rule[-1ex]{0pt}{3.5ex} \makecell{I2V Gain settings\\ } & Designed (k$\Omega$)  & Measured mean (k$\Omega$) \\
\hline
\rule[-1ex]{0pt}{3.5ex} 00 & 45\ & 39.44 \\
\hline
\rule[-1ex]{0pt}{3.5ex} 01 & 50 &  43.90 \\
\hline
\rule[-1ex]{0pt}{3.5ex} 10 & 55 &  48.33  \\
\hline
\rule[-1ex]{0pt}{3.5ex} 11 & 60 & 52.65  \\
\hline
\end{tabular}
\end{center}
\end{table}

\begin{figure} [ht]
    \begin{center}
    \begin{tabular}{c} 
    \includegraphics[width=0.6\textwidth]{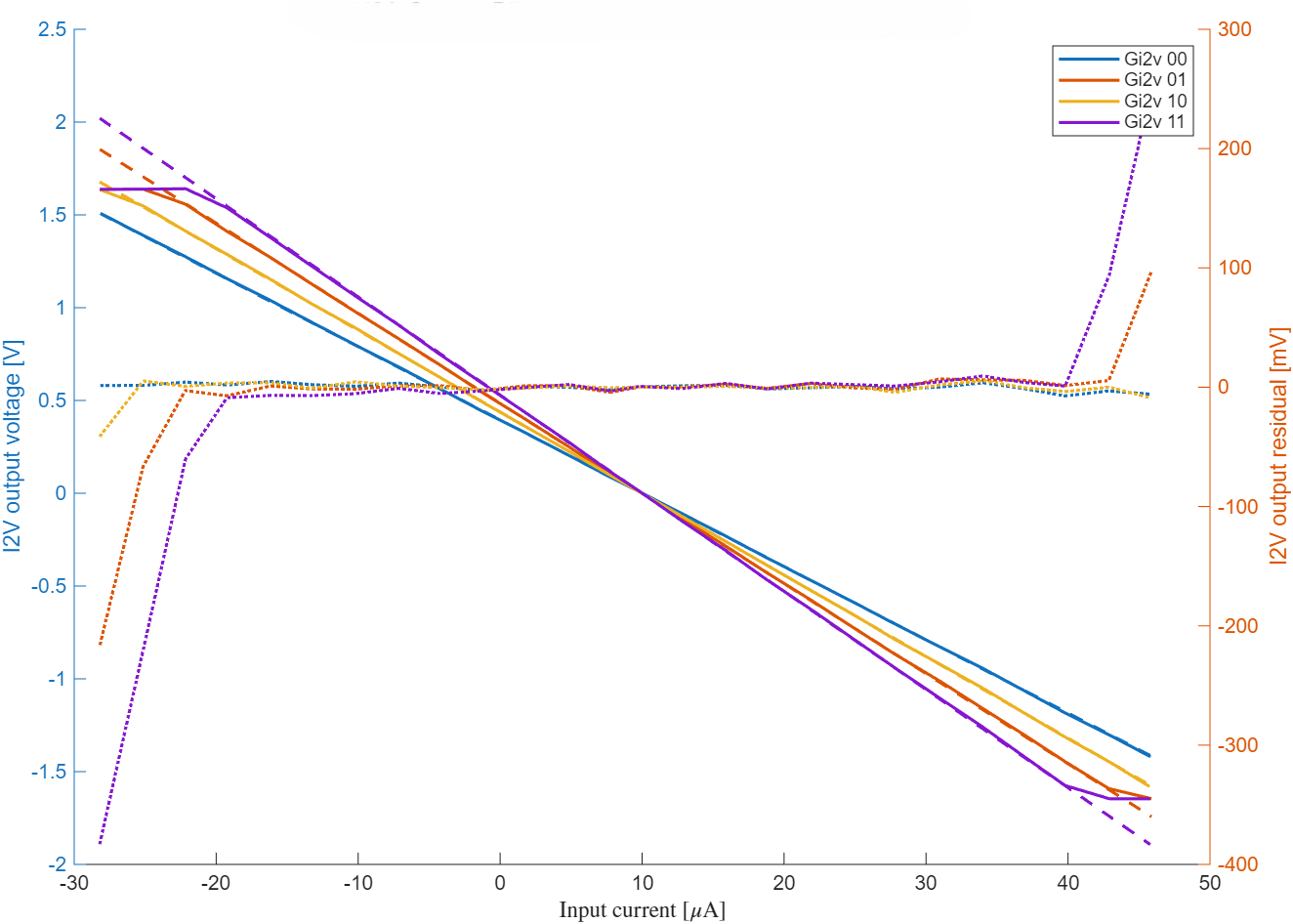}
    \end{tabular}
    \end{center}
   \caption[example] 
   { \label{fig:I2V_2} 
    Measured output voltage of the I2V as a function of input current for all 4 gain settings. The input dynamic range extends from approximately 60 µA at the higher gain setting to approximately 80 µA at the lower gain setting.\\}
\end{figure}

\begin{figure} [ht]
\begin{center}
\begin{tabular}{c} 
\includegraphics[width=0.65\textwidth]{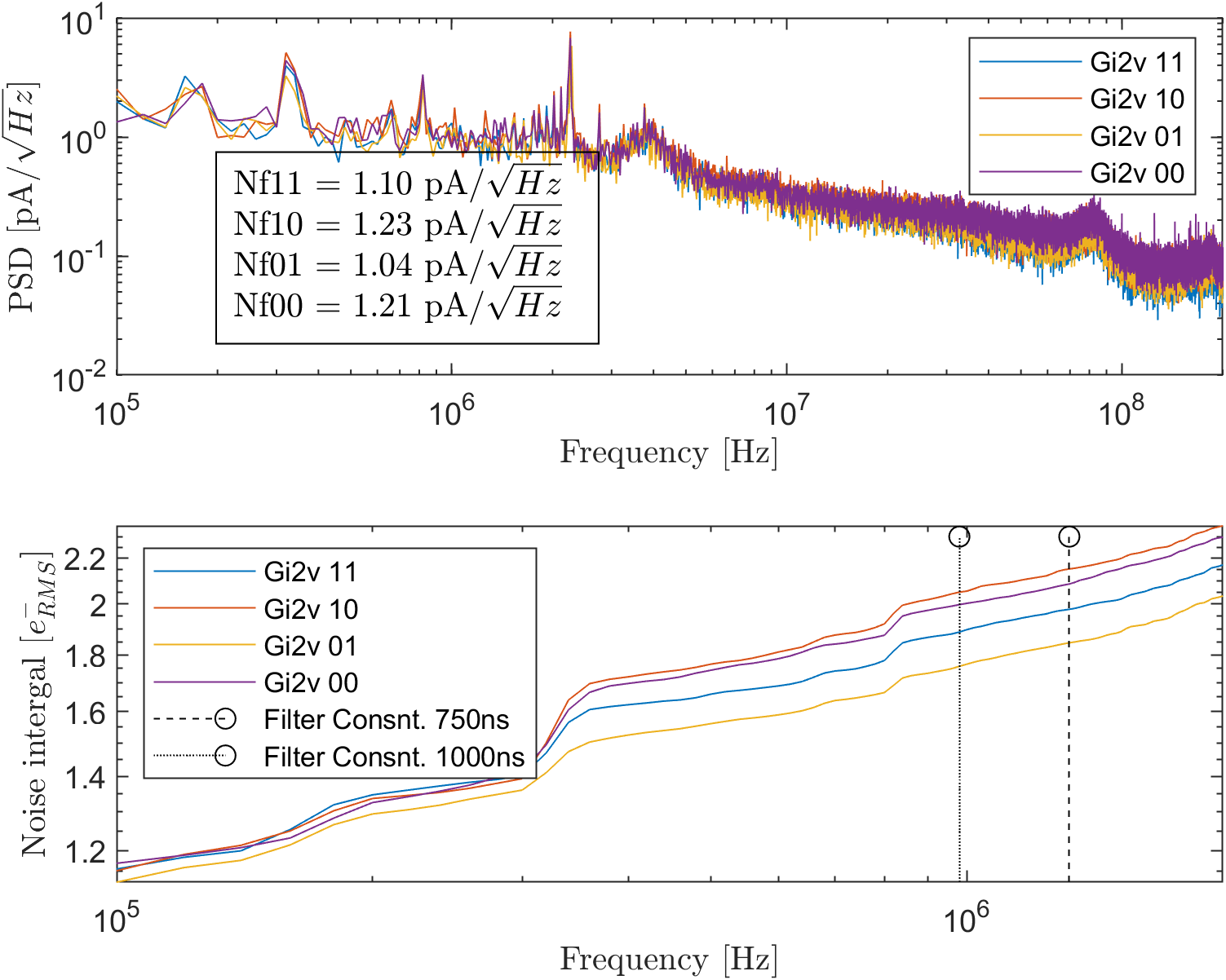}
\end{tabular}
\end{center}
\caption[example] 
{ \label{fig:I2V_3} 
I2V noise spectrum (top) and the corresponding noise integral (bottom) for all four gain settings.}
\end{figure}

\subsection{Preamplifier gain}

The preamplifier gain was characterised using two measurement configurations. In the first configuration, the multiplexer (MUX) was disabled, allowing both the preamplifier and output buffer to fully settle before the output was measured. In the second configuration, the MUX was operated under nominal conditions with 50 ns switching interval, enabling evaluation of the impact of a finite settling time on the signal amplitude.\cite{porro2014veritas, herrmann2018veritas, anna2024}

\begin{figure} [ht]
\begin{center}
\begin{tabular}{c} 
\includegraphics[width=0.8\textwidth]{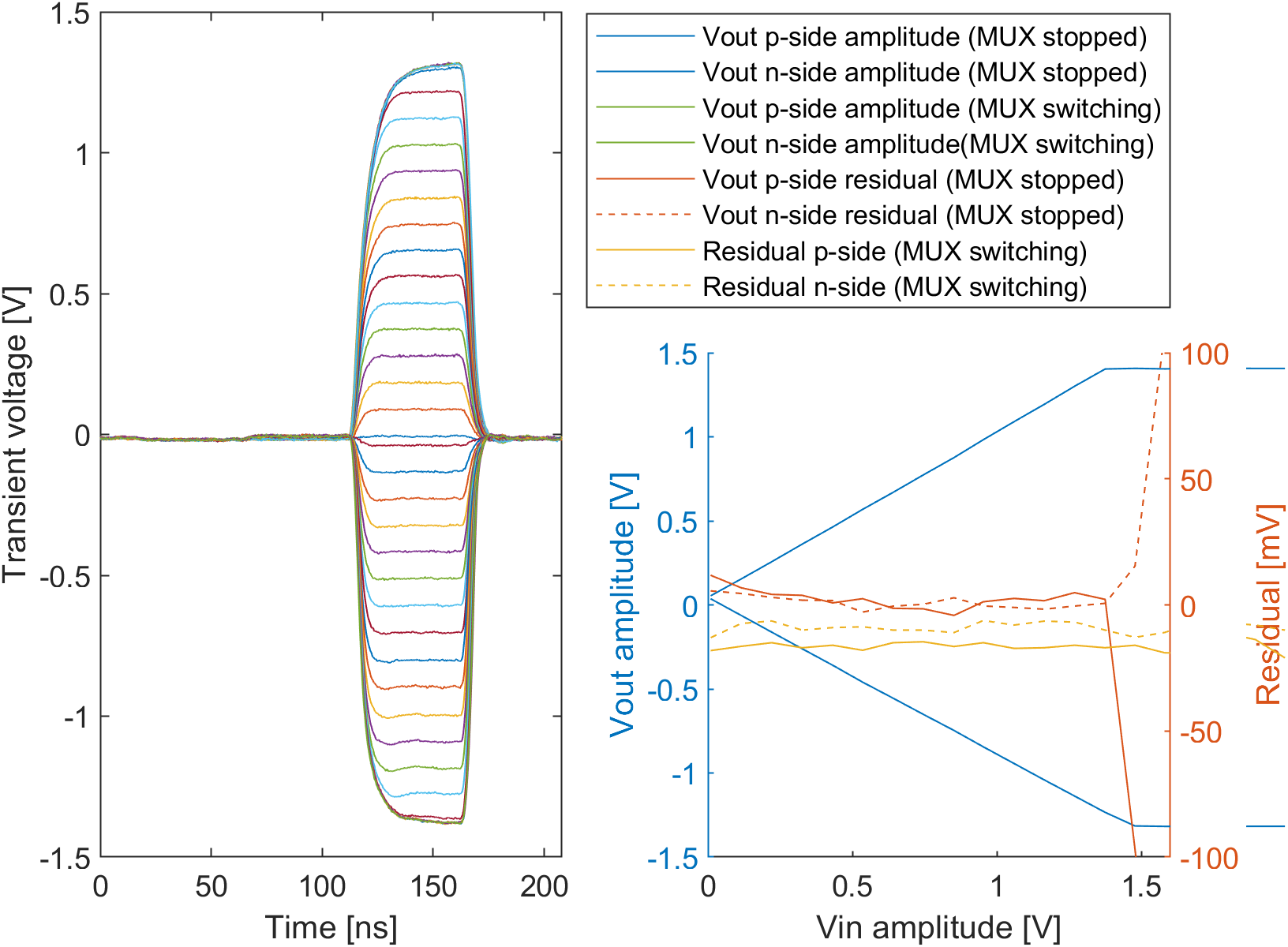}
\end{tabular}
\end{center}
\caption[example] 
{ \label{fig:preamp_1} 
The preamplifier gain has been characterised for a gain setting of 2 V/V. The left-hand plot shows the transient response under input voltage amplitude sweeps. The sampled output amplitude for both the p-side and n-side channels is presented as a function of the applied input voltage amplitude, demonstrating the corresponding gain behaviour and symmetry of the response across both polarities.}
\end{figure}

Figure~\ref{fig:preamp_1} compares the measured responses for both configurations at a preamplifier gain setting of 2 V/V. Good agreement is observed at lower signal amplitudes, whereas deviations become apparent at higher amplitudes, where the multiplexed signal exhibits incomplete settling. These results highlight the influence of the available settling time on the achievable signal accuracy under dynamic operating conditions. The measurements were repeated for all available preamplifier gain settings (4, 8, and 16 V/V), and the corresponding results are summarised in table~\ref{tab:Preamp_2}.

\begin{table}[ht]
\caption{Preamplifier measured gain.} 
\label{tab:Preamp_2}
\begin{center}       
\begin{tabular}{|c|c|c|c|c|} 
\hline
\rule[-1ex]{0pt}{3.5ex} \makecell{Preamplifier Gain settings [V/V]\\} & Measured mean gain [V/V]  \\
\hline
\rule[-1ex]{0pt}{3.5ex} 2 & 1.89 \\
\hline
\rule[-1ex]{0pt}{3.5ex} 4 & 3.72 \\
\hline
\rule[-1ex]{0pt}{3.5ex} 8 & 7.29 \\
\hline
\rule[-1ex]{0pt}{3.5ex} 16 & 13.46  \\
\hline
\end{tabular}
\end{center}
\end{table}

\subsection{Filter time constant}
The functionality and temporal response of the filter stage were evaluated by injecting a voltage step into the VERITAS 2.3.1 input, emulating the detector signal generated during charge removal. The stimulus was generated by a waveform generator, and the filter output was monitored with an oscilloscope. To enable direct observation of the filter response, the MUX/S\&H stage was bypassed, allowing the filter output to be continuously buffered and routed to the output buffer.

\begin{figure} [H]
\begin{center}
\begin{tabular}{c} 
    \includegraphics[width=0.7\textwidth]{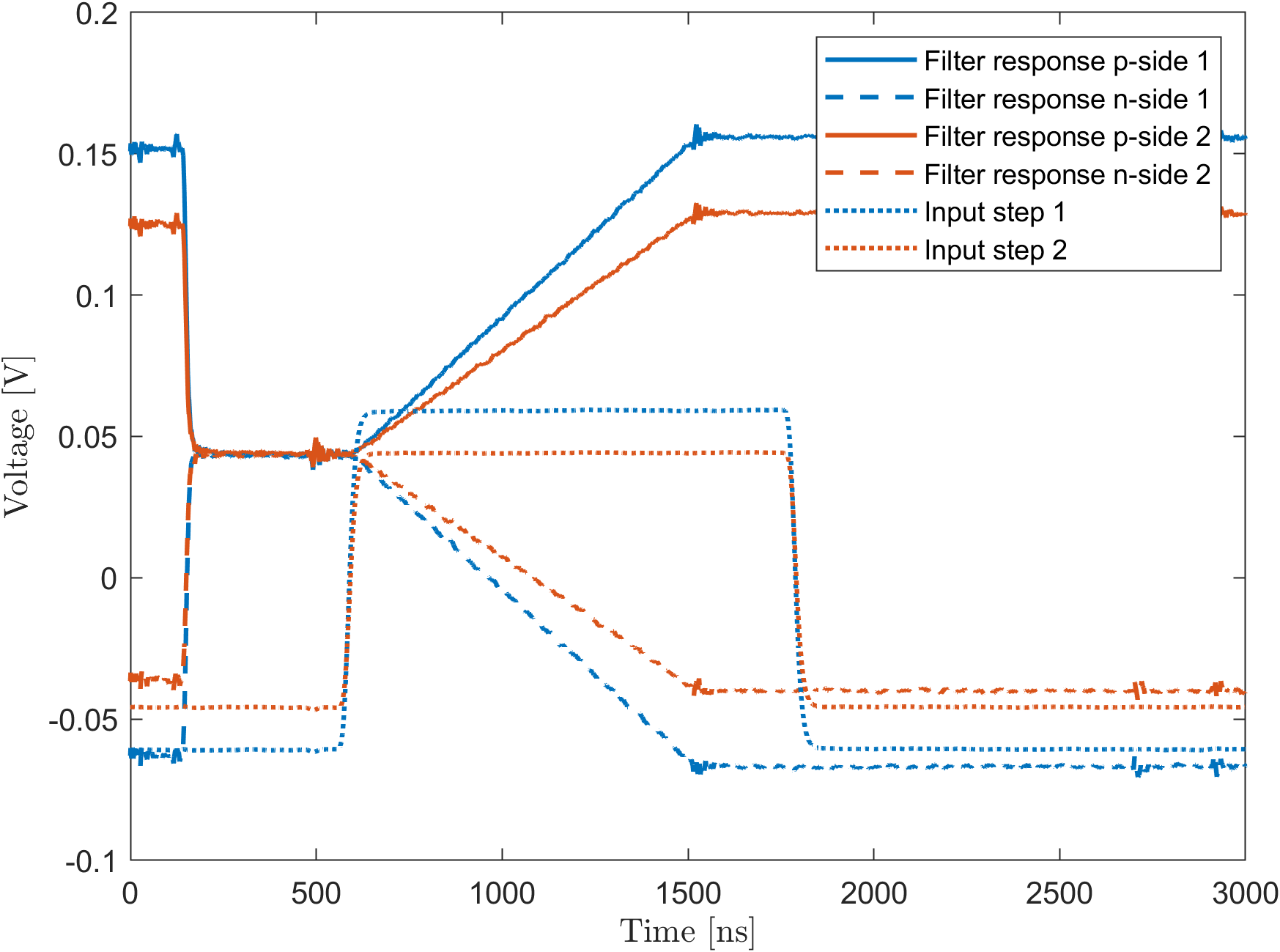}
    
\end{tabular}
\end{center}
\caption[example] 
    { \label{fig:filter} 
    Filter transfer function with positive integration. For filter gain = 0, Measured time constant = 708.2 ns and or filter gain = 1, Measured time constant = 1062.8 ns.
 }
\end{figure}

The Correlated Double Sampling (CDS) filter incorporates two user-selectable time constants as shown in figure~\ref{fig:filter} defined by programmable filter-gain resistors and feedback capacitors. By varying the resistor-capacitor combinations, the filter characteristics can be adjusted to accommodate different detector operating conditions and signal requirements. The measurements were performed to verify the implemented time constants, evaluate the filter's dynamic behaviour, and confirm correct operation across the available configuration settings.\cite{astrid2024, porro2014veritas, herrmann2018veritas, anna2024}

\subsection{Output buffer and Debugging buffer transient response}
\begin{figure} [H]
\begin{center}
\begin{tabular}{c} 
    \includegraphics[width=0.8\textwidth]{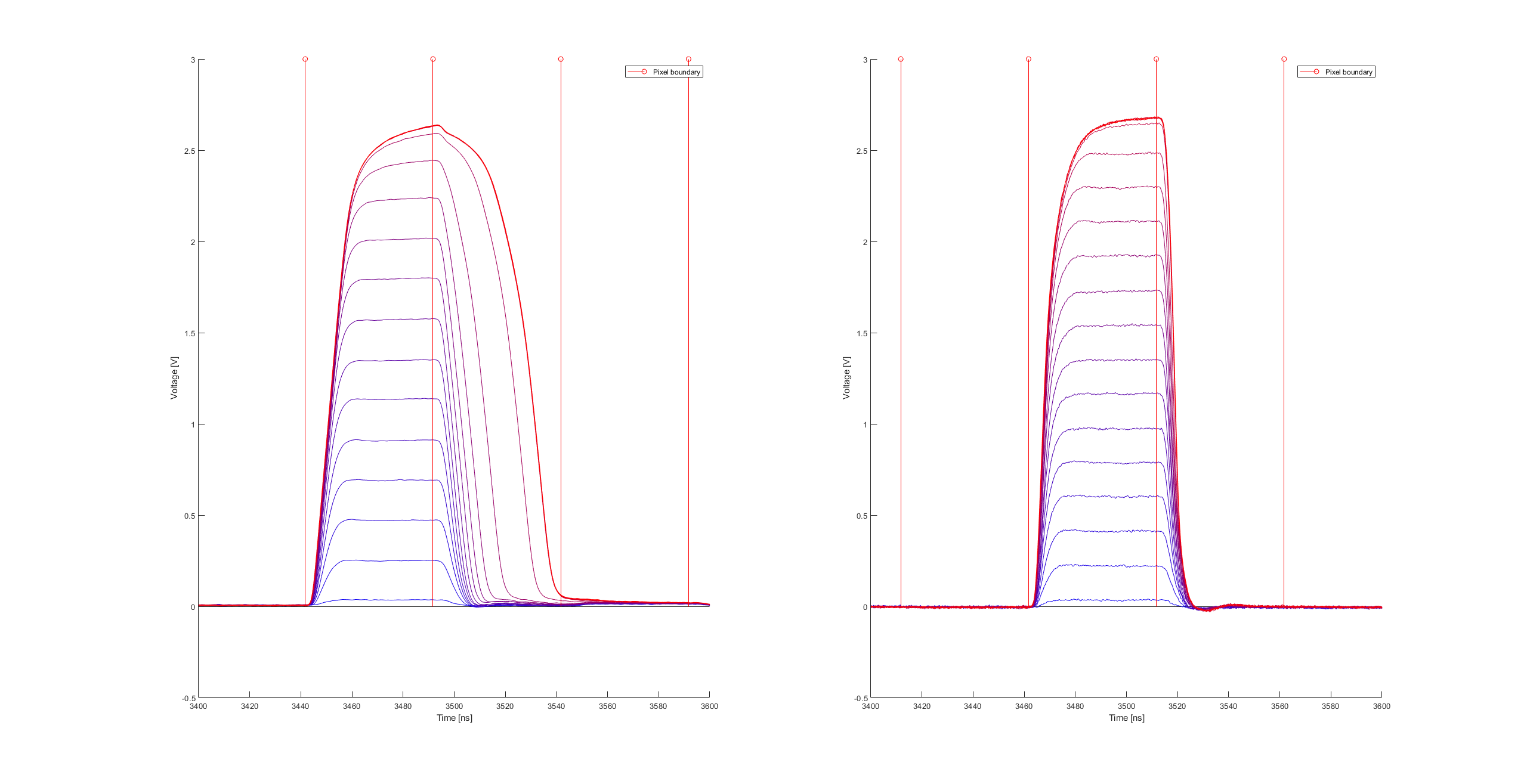}
    
\end{tabular}
\end{center}
\caption[example] 
    { \label{fig:Buf} 
    Measured transient responses of the output buffer (left) and debugging buffer (right).}
\end{figure}

Figure~\ref{fig:Buf} presents the transient responses of the output buffer and debugging buffer, respectively. The measurements show that the output buffer exhibits settling issues when driven into overdrive conditions. Although such behaviour is not expected during normal operation, it may occur when a detector pixel saturates, resulting in excessive signal levels at the input stage and, consequently, overdriving the output buffer. In this scenario, the resulting transient artefacts may extend into subsequent readout cycles, potentially corrupting the measurement of adjacent pixels.

To mitigate this effect, the operating current of the output buffer was reduced from approximately 150 $\mu$A to 50 $\mu$A. This optimisation significantly reduces the transient overshoot and crossover behaviour observed under overdrive conditions, resulting in improved settling characteristics and enhanced robustness against pixel saturation events.\cite{herrmann2018veritas, anna2024}

\subsection{Cross talk}

\begin{figure}[H]
\begin{center}

\begin{tabular}{c}

    \includegraphics[width=0.5\textwidth]{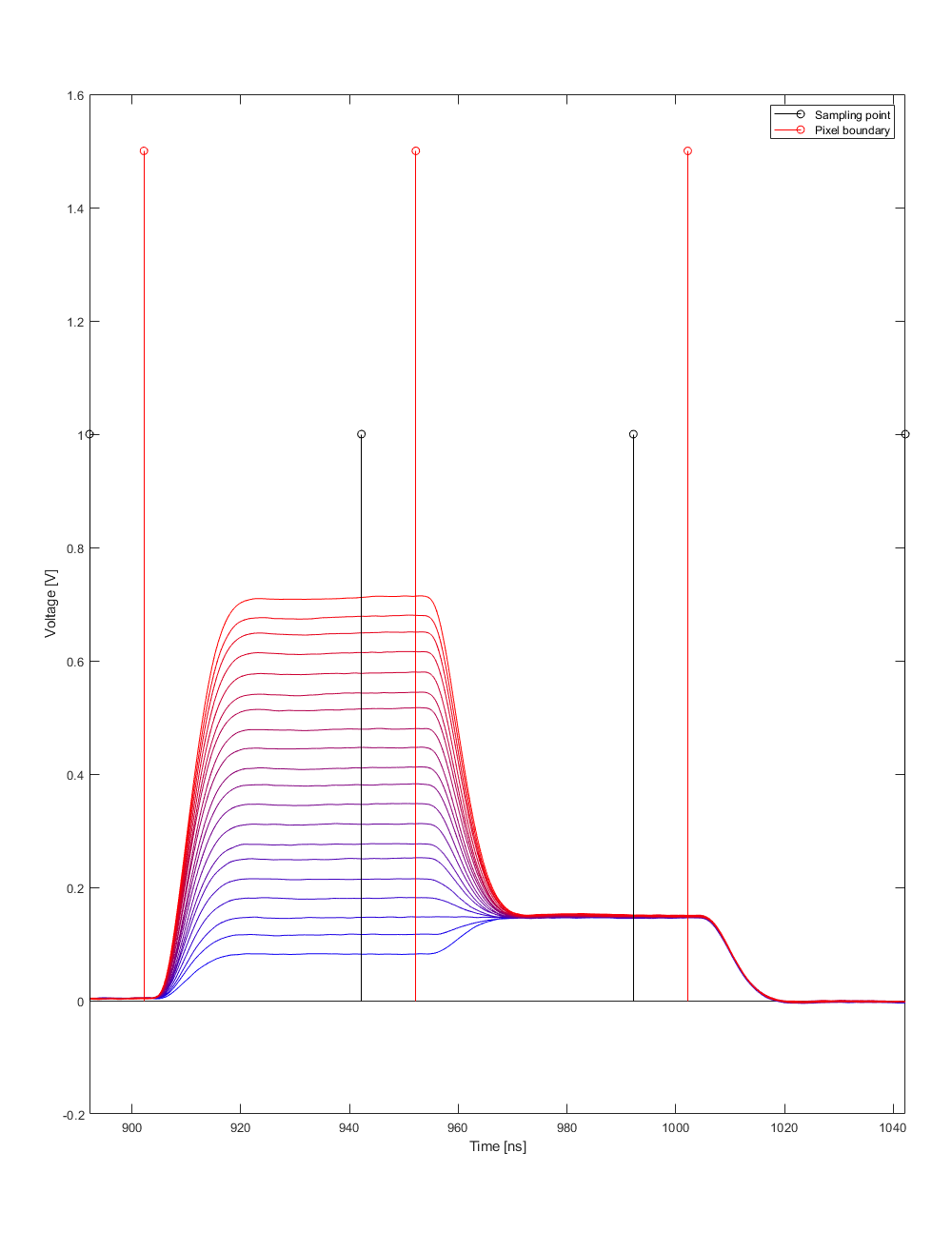}

\end{tabular}
\end{center}
\caption[example]
{\label{fig:cross}
Crosstalk measurement between adjacent channels, with the adjacent channel driven by a pulse. The new output stage fixes the adjacent channel cross-talk compared to the previous chip iteration.}
\end{figure}

The cross talk measurements were performed using the output buffer configuration as shown in figure~\ref{fig:cross}, with the adjacent channel driven by a fixed-amplitude pulse. Under driven conditions, the isolation is approximately 49 dB, equivalent to an attenuation factor of around 281. The results demonstrate that the coupling between adjacent channels is significantly reduced when the neighbouring channel is actively driven, indicating improved immunity to cross talk under normal operating conditions.\\

\subsection{Performance measurement with DEPFET}

\begin{figure} [H]
\begin{center}
\begin{tabular}{c} 
    \includegraphics[width=0.5\textwidth]{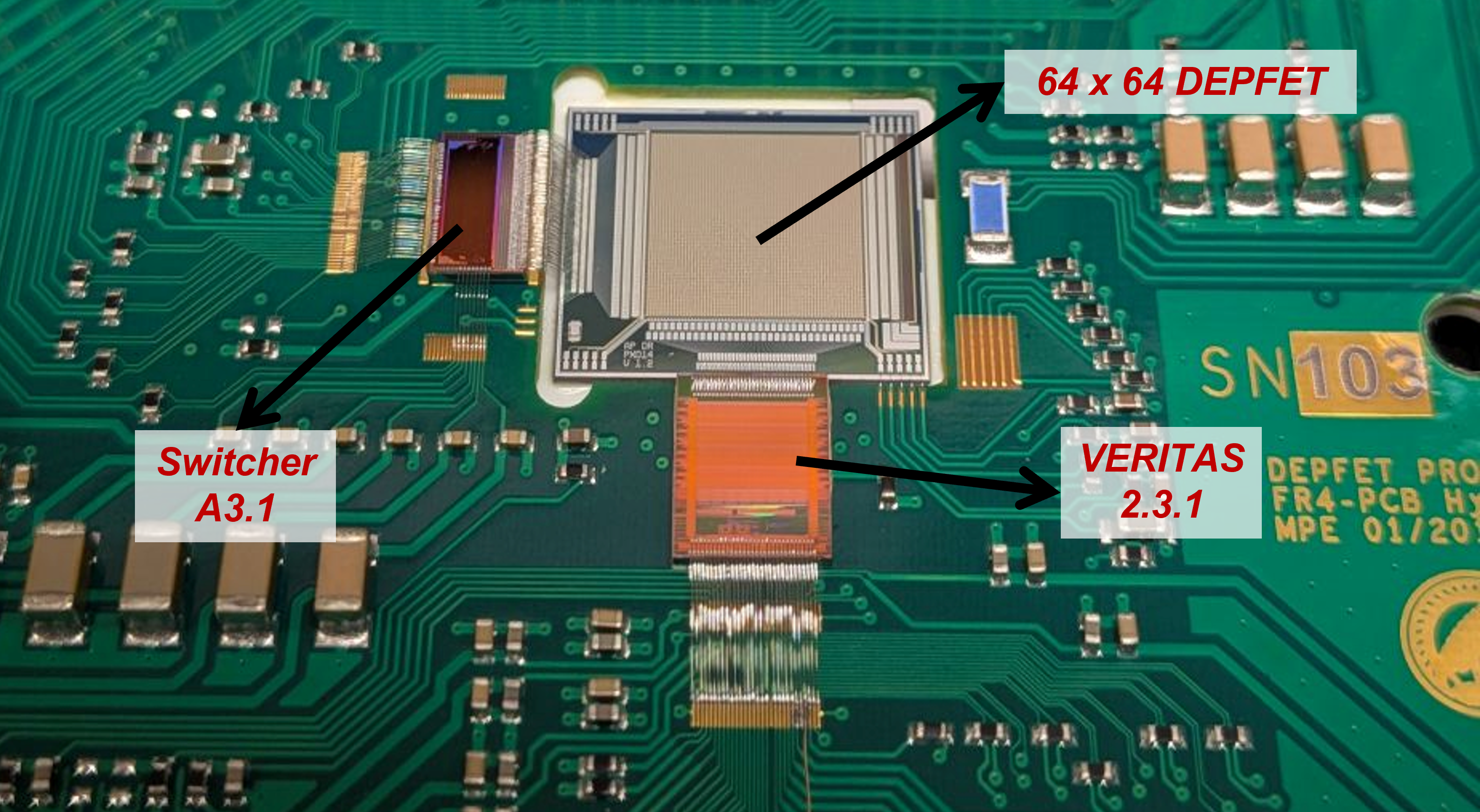}
    
\end{tabular}
\end{center}
\caption[example] 
    { \label{fig:spectrum_1} 
    Ceramic-hybrid PCB with 64 x 64 DEPFET and ASICs.
 }
\end{figure}

Spectroscopic characterisation\cite{treberspurg2016studies, treberspurg2017studies, bonholzer2018first, bonholzer2022drain} was performed using a single-event spectrum acquired from a $^{55}\mathrm{Fe}$ source, with the ASICs coupled to a 64 × 64 DEPFET prototype operating in drain readout mode. The detector features a pixel pitch of 130 × 130 µm². Figure~\ref{fig:spectrum_1} shows the ceramic-hybrid PCB module used for the measurements, on which the VERITAS 2.3.1 and Switcher ASICs are mounted and wire-bonded to the detector.

The Switcher ASIC provides the control signals for row selection and DEPFET matrix clearing, while VERITAS 2.3.1 performs the signal readout. A spectral resolution of 130 eV FWHM was achieved at an operating temperature of $-66\,^\circ\mathrm{C}$. During these measurements, both the signal-plus-baseline and baseline acquisition phases were configured with an integration time of $1\,\mu s$. The DEPFET clear operation was performed during the flat-top interval of the acquisition cycle. The measured mean noise performance of VERITAS 2.3.1 was $\sim$ 3.1\,\(e^-\) ENC per pixel (for reference, VERITAS 2.2 was $\sim$ 2.9\,\(e^-\) ENC per pixel), demonstrating excellent low-noise operation and confirming the ASIC's suitability for high-resolution X-ray spectroscopy applications.\\

\begin{figure} [H]
\begin{center}
\begin{tabular}{c} 
    \includegraphics[width=0.9\textwidth]{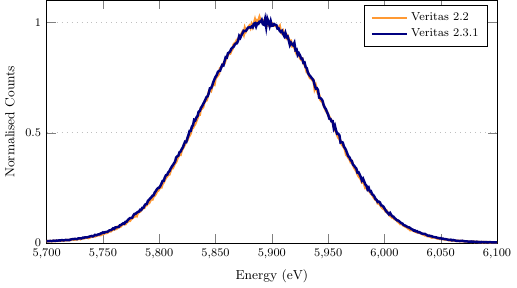}
    
\end{tabular}
\end{center}
\caption[example] 
    { \label{fig:spectrum_2} 
    Measured Spectrum of ~$^{55}\mathrm{Fe}$ source for VERITAS 2.2 (AMS) and 2.3.1 (XFAB) exhibit comparable performance, confirming the successful migration of the design.
 }
\end{figure}

\section{CONCLUSION} 
The VERITAS 2.3.1 validation successfully demonstrates that the transition to the XFAB semiconductor technology process node has also been completed. Furthermore, the new VERITAS ASIC meets all specified design requirements. A spectral performance of 130 eV FWHM has been achieved using the DEPFET drain readout architecture, demonstrating excellent suitability for high-resolution X-ray spectroscopy applications. The device has reached Technology Readiness Level 4 (TRL), with radiation qualification activities, including total ionising dose (TID) and single-event effects (SEE) testing, currently planned in order to advance the device to TRL5.

The observed reductions in I2V feedback resistance, filter time constant, and preamplifier gain are attributed to systematically lower resistance values across the fabricated chips, indicating a process-dependent variation introduced by the foundry. This behaviour is consistent with a global shift in resistor sheet values within the XFAB process rather than an isolated circuit-level anomaly.

\acknowledgments 
 
The work was funded by the German space agency DLR (FKZ: 50 QR 2301 and FKZ: 50 QR 2501) and the NASA contribution to Athena (Contract: 80GSFC21C0005). 

\bibliography{report} 
\bibliographystyle{spiebib} 

\end{document}